\documentclass[aip,rsi,reprint]{revtex4-1}
\usepackage{amssymb}
\usepackage{graphicx}
\usepackage{dcolumn}
\usepackage{amsmath}
\usepackage{float}
\input{tcilatex}
\usepackage{color}
\usepackage{ulem}

\begin{document}

\title{Highly sensitive simple homodyne phase detector for ultrasonic pulse-echo measurements}

\author{John Grossmann}
\affiliation{Colgate University, Hamilton, NY 13346}
\author{Alexey Suslov}
\affiliation{National High Magnetic Field Laboratory, Tallahassee, FL 32310}
\author{Grace Yong}
\affiliation{Towson University, Towson, MD 21252}
\author{Lynn A. Boatner}
\affiliation{Oak Ridge National Laboratory, Oak Ridge, TN 37831}
\author{Oleksiy Svitelskiy}
\email{oleksiy.svitelskiy@gordon.edu}
%\affiliation{Colgate University, Hamilton, NY 13346}
\affiliation{Gordon College, Wenham, MA 01984}
%\email{oleksiysvit@gmail.com}

\begin{abstract}
We have designed and built a modern versatile research-grade instrument for ultrasound pulse-echo probing of the elastic properties of a wide range of materials under laboratory conditions. The heart of  the instrument lies in an AD8302 microchip: a gain and phase detector from  Analog Devices, Inc. To construct the device, we have implemented a schematic that utilizes the homodyne principle for signal processing instead of the traditional superheterodyne approach. This design allows one to measure phase shifts with high precision and linearity over the entire range of $0-360^\circ$. The system is simple in construction and usage; it makes ultrasound measurements easily accessible to a broad range of researchers. It was tested by measuring the temperature dependence of the ultrasound speed and attenuation in a KTa$_{0.92}$Nb$_{0.08}$O$_{3}$ (KTN) single crystal at a frequency of $\sim$ 40 MHz. The tests were performed in the vicinity of the ferroelectric transitions where the large variations of the speed and attenuation demand a detector with outstanding characteristics. The described detector has a wide dynamic range and allows for measuring in a single run over the whole temperature range of the ferroelectric transitions, rather than just in limited intervals available previously.   Moreover, due to the wide dynamic range of the gain measurements and high sensitivity this instrument was able to reveal previously unresolvable features associated with the development of the ferroelectric transitions of KTN crystals.
\end{abstract}

\maketitle

\section{INTRODUCTION}

\begin{figure}[th]
\includegraphics[width=1\columnwidth]{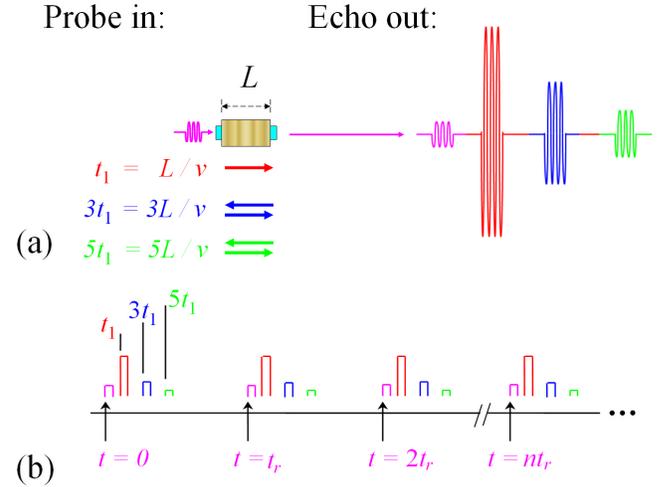}
\caption{(Color online) Idea of the ultrasound pulse-echo experiment. (a) Train of echoes on the output transducer following the input RF pulse. (b) Sequence of repetitive pulse-echo signals: envelops of the RF pulses are presented.}
\label{fig1}
\end{figure}

Subsequent to its development more than fifty years ago\cite{McSkimin,Viswanathan}, the ultrasonic pulse-echo technique has proven to be a valuable and indispensable non-destructive tool for exploring elastic properties of materials that are pertinent to both applied and basic investigations. In this technique, an electric radiofrequency (RF) pulse is applied to a transducer that is firmly affixed to the sample under investigation (Fig.\ref{fig1}). The transducer produces an acoustic pulse that travels through the sample to another transducer where a small part of the acoustic energy is converted into an electric signal, while the remaining portion is reflected back into the sample (i.e., towards the first transducer), where it essentially ``echoes'' forward, and so on. Thus, one acoustic pulse travels back and forth through the sample many times, and therefore, a single probe pulse causes a train of echoes as shown in Figure \ref{fig1}(a). For data acquisition feasibility and for signal averaging, the probe pulse is repeated with a periodicity $t_{r}$ that results in a sequence of pulse and echo signals (see Fig.\ref{fig1}(b)).  Knowing the length of the sample $L$ it is then straightforward to determine the speed of the corresponding type of acoustic wave $v=(2n-1)L/t_{n}$, where $n$ is the echo number and $t_{n}$ is the time delay between the probe and this echo. Measuring the speed of sound waves with different polarizations and by propagating the pulses along various crystallographic directions of the sample, enables one to solve the system of Christoffel equations\cite{Auld}, and thereby, determine the full set of the elastic parameters of the material. At the same time, measurements of the signal amplitude allow one to study the mechanisms of sound attenuation in the material. 

In order to increase the precision of the measurements, a phase-detection method that compares relative amplitudes and phases of the echo and probe signals is utilized \cite{Ritec,suslov}. This method is especially useful for sensing small changes in the sound velocity when the sample is subjected to external factors, such as changing temperature, pressure, applied electric or magnetic fields, etc. 

\begin{figure*}[th]
\includegraphics[width=2\columnwidth]{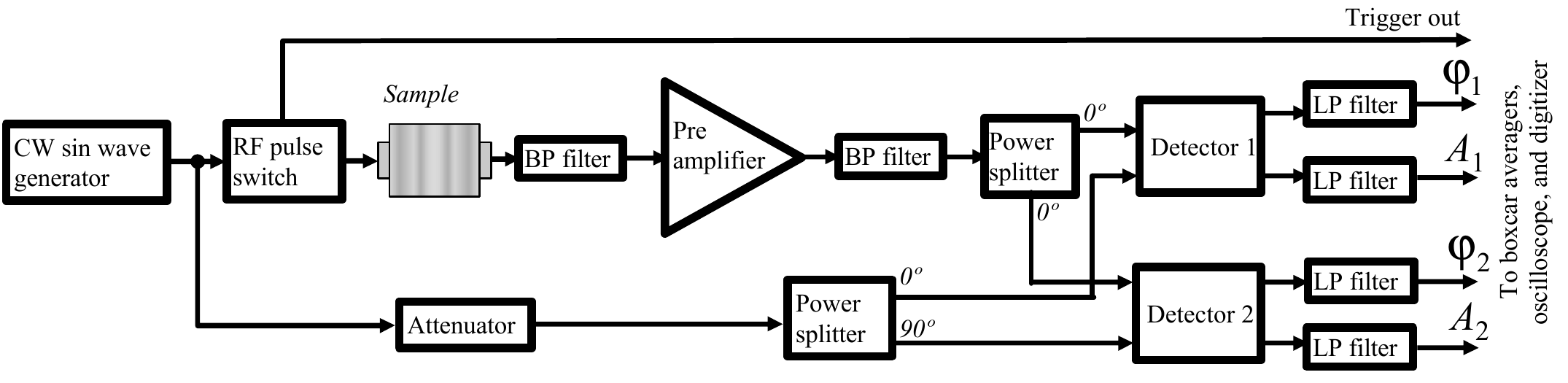}
\caption{Block diagram of the ultrasound setup. }
\label{fig2}
\end{figure*}

Almost every ultrasonic setup uses a classical phase-sensitive superheterodyne receiver, because it provides the capability for precise measurements of the ultrasound attenuation and phase over a broad frequency range. At the same time, the superheterodyne technique requires extensive circuitry for preliminary treatment of the signal before the demodulator. The design of such complicated schematics requires specialized expertize in electrical engineering, and also, its implementation is rather time and effort consuming  due to the large number of the included circuit elements. Many of these elements have to be specially selected, and the assembled circuits need tedious fine-tuning.

Progress in microelectronic technology has allowed us to significantly simplify the instrument by utilizing an AD8302 gain and phase detector microchip from Analog Device Inc\cite{Analog}. The wide frequency range of this microchip eliminates the necessity for utilizing superheterodyne circuitry and permits the processing of the signal directly at the frequency of the measurements. In some experiments, its high sensitivity and broad dynamic range allow for direct input of the transducer signal to the microchip without the use of preliminary amplification. Additionally, the linearity of the phase and gain outputs simplifies processing of the measured data. As shown below, testing of this setup demonstrated its superior phase and amplitude characteristics. The excellent characteristics of the instrument along with the simplicity of its design and usage, significantly simplifies the task of ultrasound measurements and facilitates the availability of the method for broader range of researchers and tasks.

\section{The instrument}
\label{partTheInstrument}

A block-diagram of the instrument is shown in Figure \ref{fig2}. A homodyne receiver that consists of two AD8302 detectors is the key element of our device. 
In many applications only a small change of the ultrasound velocity is observed, and consequently, measurements of a small phase variation are needed. In such settings, only one AD8302 detector would be sufficient, but the limited linear range of the chip of about $140^\circ$ would restrict the domain of the measurable sound speed variations. 
We present here a more general implementation
suitable for measurements of arbitrarily large changes in the signal phase and consequently arbitrarily large changes in the sound velocity that, for example, accompany phase transitions in solids.

In addition to a receiver, our setup also includes a generator module, which produces an ultrasound probe pulse. It forms short ($\sim 1~ \mu$s) RF pulses from the continuous signal created by a high-quality sine wave generator. This circuit also supplies a reference signal for the detectors, and trigger signals for both an oscilloscope and a boxcar averaging data processing system. Generally speaking, the generator module is independent of the receiver and can be replaced by another unit capable of producing an appropriate high-phase-stability RF burst, a related trigger, and an RF reference signal.
Below we present a detailed description of our instrument.

\subsection{Receiver}

\subsubsection{Amplitude and phase detector}

\begin{figure}[th]
\includegraphics[width=1\columnwidth]{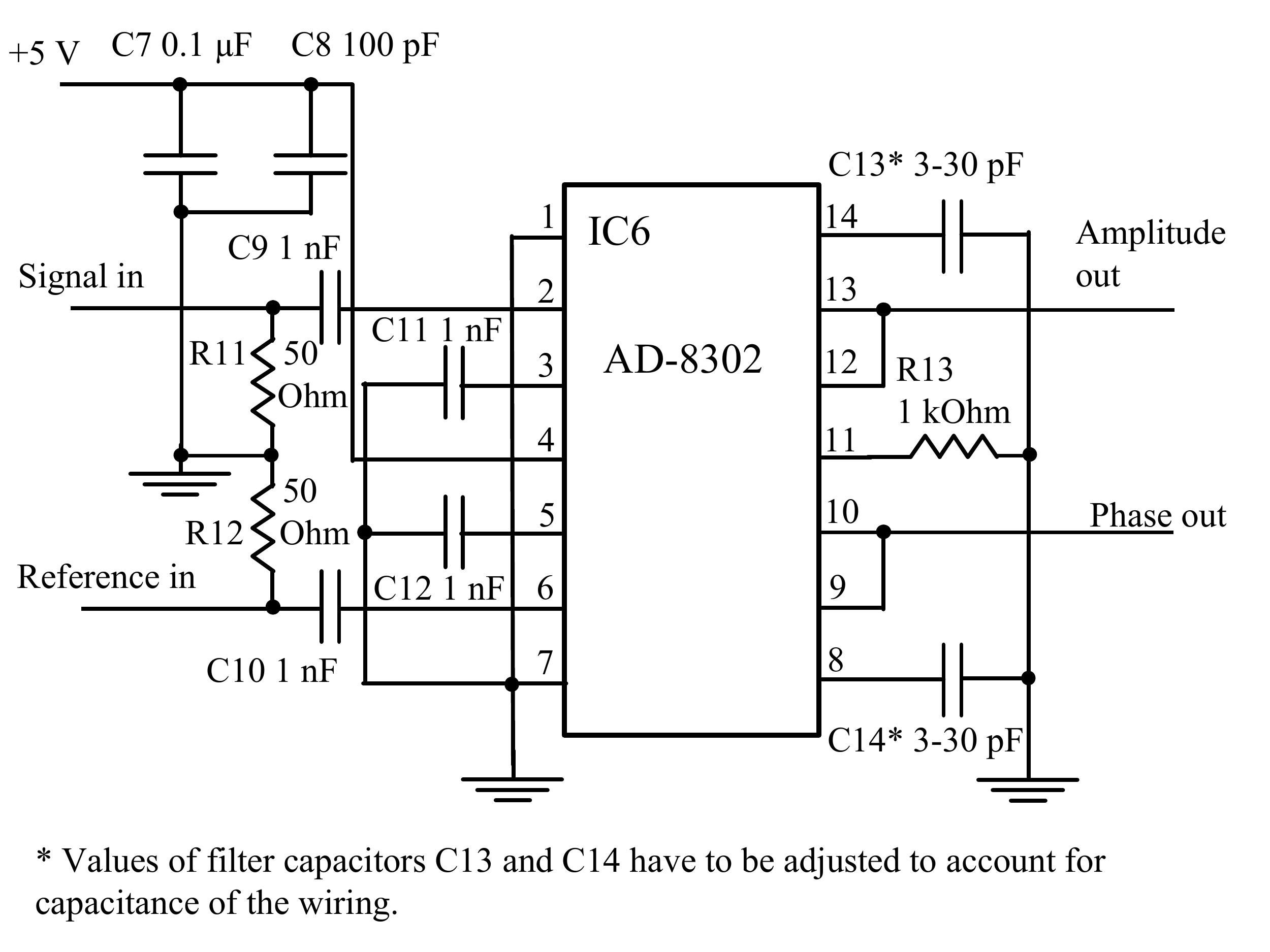}
\caption{Electrical schematic of the amplitude and phase detector.}
\label{Fig3}
\end{figure}

 Two identical detectors (Detector 1 and Detector 2 in Figure \ref{fig2}) are made on the basis of the microchip AD8302. This microchip represents a fully integrated RF system that requires only a few external components for matching the input and output impedances and for signal conditioning. This chip is capable of measuring the gain and phase of the signal with respect to the reference one. According to the manufacturer's specifications, this microchip can work at frequencies up to 2.7 GHz. It possesses a rather broad dynamic range ($\pm$30~dBm with respect to the reference signal), and a high (-60~dBm) sensitivity. The significant envelope bandwidth spans from DC to 30 MHz and allows for detecting pulses of about 1~$\mu$s as required in ultrasonic methods. 

It is possible to purchase this microchip either mounted on the evaluation board that is ready for use, or alone. Whereas the first option eliminates the hassle of making a rather difficult PCB board, the second option can be useful for student instruction. Thus, in our case, the microchips had been initially purchased alone and were mounted according to the circuit shown in Figure \ref{Fig3}. Attention has to be paid to having all of the wiring as short as possible. The power filter capacitor C8 has to be mounted as close to the microchip as possible. Capacitors C13 and C14 serve as output filter capacitors. Their values have to be adjusted in the range of 3-30~pF, depending on the capacitance of the printed circuit wiring. Many microchips can work without these capacitors as well. 

Taking at the two inputs two RF signals of the same frequency each detector produces two outputs. The first output 
is a voltage that is proportional to the logarithm of the ratio of the amplitudes of the inputs. The second (and the most important for us) output is a voltage that is linearly proportional to the phase shift between the RF input signals in the range of about 20$^{\circ}$-160$^{\circ}$ (and 200$^{\circ}$-340$^{\circ}$). Thus, the phase output produces a voltage that is linearly proportional to the phase difference between the received and the reference signals everywhere except for the relatively small non-linear regions near $0{^\circ}$ and $180{^\circ}$. These areas of nonlinearity are about $\pm 20{^\circ}$ and their exact values depend slightly on the signal frequency.

As noted above, in its minimal configuration, the receiving and processing circuit can be limited to only one gain and phase detector AD8302 (see Appendix A), whose two inputs are fed with the received signal and the reference signal, respectively. However, 
our circuit is equipped with a second AD8302 detector and a number of other concomitant elements. This allows for expanding the range of the phase measurement capabilities over the entire $360^\circ$, i.e. outside the linear range of $140^\circ$ characteristic of a single microchip. 

In our schematics, the received signal (after passing through a preamplifier, a filter, and a two-way  $0^{\circ}$ power splitter) goes to the first inputs of two AD8302 detectors. The second inputs of these detectors are fed with the reference signal that is coming from the CW sine wave generator through an attenuator and a two-way $0^{\circ}$-$90^{\circ}$ power splitter. The employment of two detectors excludes the above-noted nonlinearities in the phase characteristics: whenever one of the phase detectors is in the nonlinear regime, the other detector gives reliable and easy to interpret linear readings as we show below in the test section. The attenuator is necessary to ensure that the levels of both input signals are optimized in the working range of the  AD8302 detectors. Bandpass and lowpass filters (labeled as BP filter and LP filter in Figure \ref{fig2}, respectively) were used for reducing the noise outside of the frequency range of interest, i.e., for increasing the signal-to-noise ratio.
All phase and amplitude voltages are further processed with SR250 boxcar averagers \cite{SRS} in order to concentrate the analysis on particular echoes. The boxcars are triggered by the trigger pulses from the RF pulse switch. In our tests, we used the repetition frequency of 10~kHz and averaged the signal over 3~k samples. Such parameters specify reasonably small time constant of about 0.3~s and, at the same time, allow for a significant increase in the signal to noise ratio. The outputs of the boxcars through the National Instruments PCI-5105 digitizer\cite{NI} are input for further processing  into a computer. 

\subsubsection{Preamplifier}

\begin{figure}[th]
\includegraphics[width=1\columnwidth]{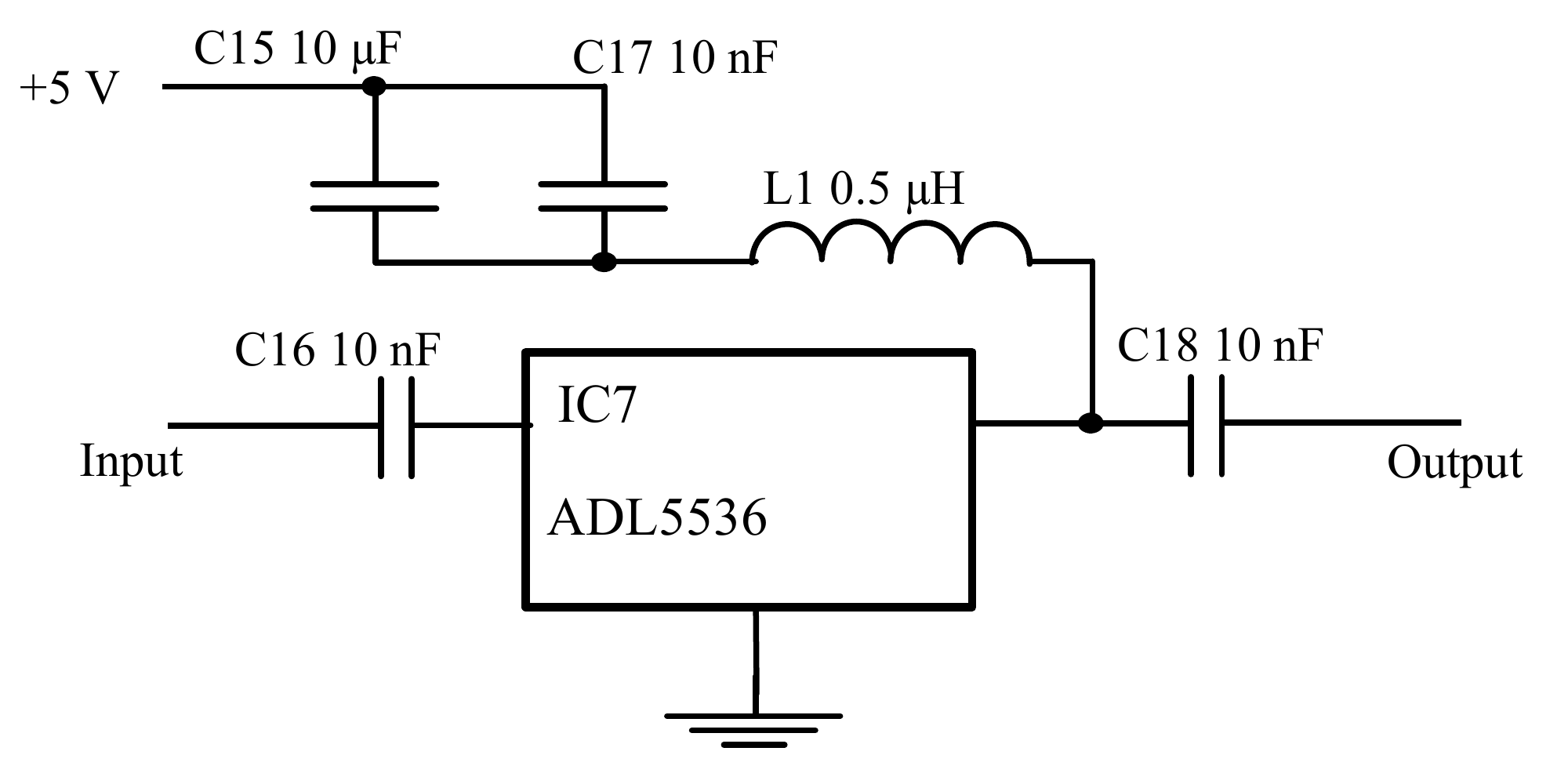}
\caption{Electrical schematic of the preliminary amplifier.}
\label{Fig4}
\end{figure}

Although the sensitivity of the RF inputs of the AD8302 chip is sufficiently high to satisfy the needs of many experimentalists, certain applications may require a preliminary amplification circuit. We have built such a preliminary amplifier on the microchip ADL5536\cite{Analog}, as shown in Figure \ref{Fig4}. This microchip has a fixed gain of 20~dB and can operate in the frequency range of 20 - 1000~MHz. While mounting this microchip, it is important to provide RF isolation between its input and output circuits in order to prevent self-excitation at high frequencies. Alternatively, one can use a factory-made amplifier, for example, ZFL-100LN from Mini-Circuits, Inc\cite{Minicircuits}. The characteristics of this amplifier are comparable to ours, however, it requires a 15~V power source. 

\subsubsection{Other moduli of the receiver}

The attenuator was made using resistors arranged in standard schematics that match the necessary input and output impedance values. The two-way $0^{\circ}$ power splitter is made of three 50~Ohm resistors connected in a triangle. The two-way $90^{\circ}$ power splitter was a ZMSCQ-2-50 device from Mini-Circuits, Inc.\cite{Minicircuits}. This power splitter guarantees a phase shift of 90$^{\circ}\pm3^{\circ}$ with an amplitude unbalance not exceeding 1.5~dB in the frequency range of $25-50$~MHz. To prevent parasitic crosstalk between different moduli, each module is screened by placing it in a separate metal box. The bandpass filters were combined  by connecting lowpass BLP-50 and highpass BHP-50 filters from Mini-Circuits, Inc.\cite{Minicircuits}. The lowpass filters SLP-10 from the same company were used at the detector outputs. It should be noted that the models of the power splitter and the bandpass filters mentioned above were specifically selected for matching the frequency of interest, namely the main frequency of the piezoelectric transducers used in the tests. These models restrict the operating frequency of the receiver to the range from about 37~MHz to about 55~MHz. At the same time the bandwidth of the AD8302 detector extends up to 2.7 GHz. Therefore, these power splitters and the filters should be replaced by other units if measurements on a frequency lying out of the range of 37~MHz --- 55~MHz are required. Appropriate modules are commercially available. We tested our detector in the frequency range 25~MHz --- 550~MHz, which is the operating bandwidth of our transducers.

\subsection{Producing RF probe pulses}

\begin{figure}[th]
\includegraphics[width=1\columnwidth]{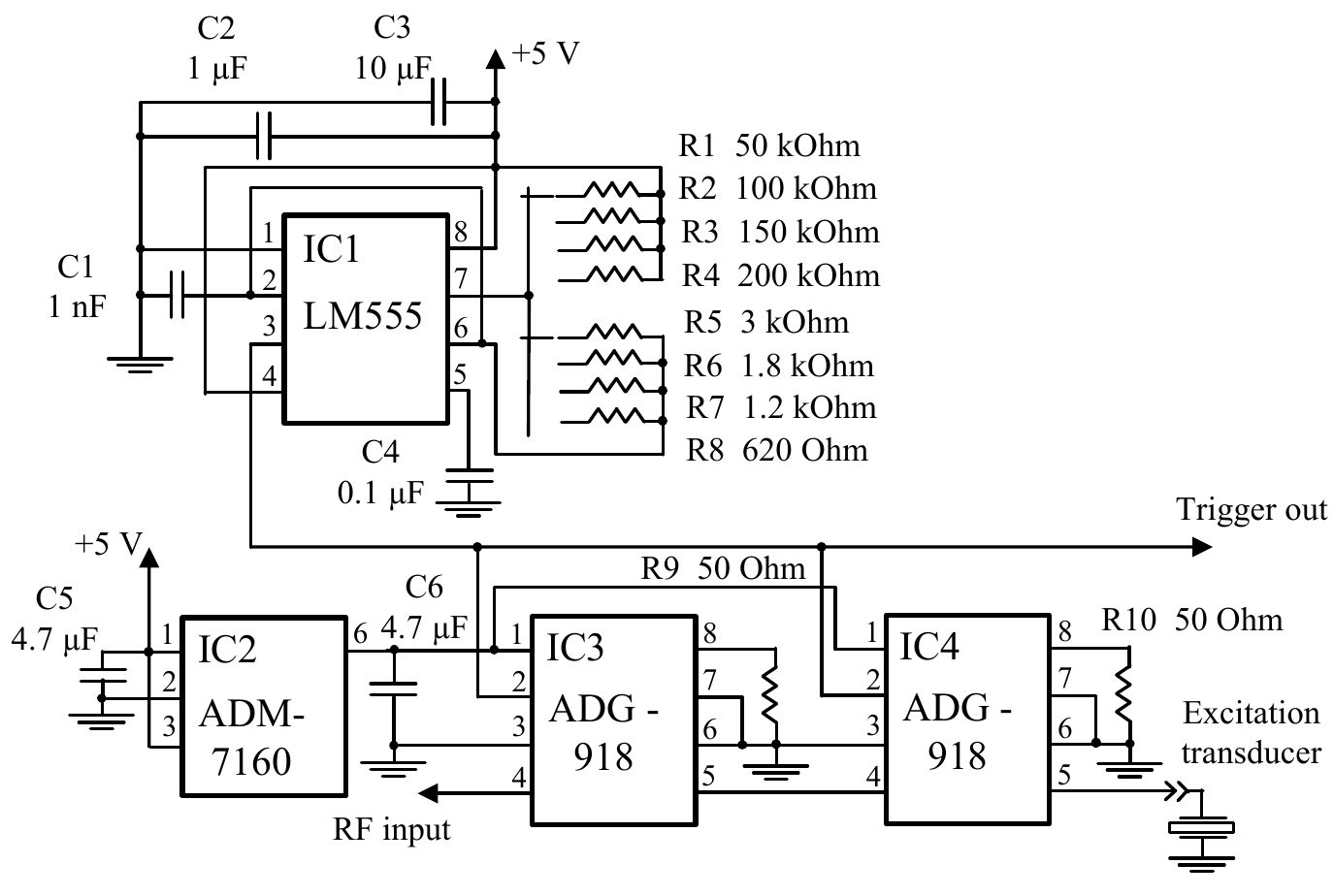}
\caption{Electrical schematic of the circuit producing the probe RF pulses.}
\label{fig5}
\end{figure}

The probe pulses are produced from the harmonic signal of an Agilent 33250A function generator \cite{Agilent} by means of an RF switch that consists of a series of two ADG918 absorptive microswitches\cite{Analog}  (Fig. \ref{fig5}). The switch is controlled by rectangular pulses from the LM555 timer\cite{Texas}. The length of these pulses can be chosen between $1~\mu$s, $1.5~\mu$s, $2~\mu$s, and $3~\mu$s; the pulse repetition rate can be 5~kHz, 6.7~kHz, 10~kHz, or 20~kHz.  The microchip ADM7160\cite{Analog} is a voltage regulator that produces a stable voltage of +2.5~V that is necessary to feed the ADG918s.  With such a switch one can produce nicely shaped rectangular RF pulses up to 2~V in amplitude with a -40~dB suppression of the RF signal in between the pulses. In all our tests described below, we used RF bursts with an amplitude of about 0.5~V, which was sufficient for obtaining a good quality ultrasound signal. For the best result, it is important to use a highly coherent and stable function generator. For instance, the above-mentioned Agilent 33250A generator, according to the manufacturer's specifications, has a frequency stability of 2 ppm/year in the temperature range between 18 and 28$^{\circ}$C.

\section{Results of testing}
\label{partTest}

Interest in the physical acoustics associated with the phase transitions originates, in particular, from the fact that measurements of the velocity variation at a transition allow for calculating the change of the elastic moduli and, therefore, finding the coefficients of the free energy expansion in the order parameter \cite{Landau, Rehwald}. However, it is well known that in the vicinity of phase transitions a significant rise of the ultrasound attenuation and a considerable change of ultrasound velocity are observed. This leads to substantial experimental difficulties in acoustic studies of these phenomena. Therefore, the acoustic measurements near the phase transitions require detectors which have high sensitivity, large dynamic range for gain measurements, and  that are capable of measuring the phase change over the entire $0-360^\circ$ region. 

At the same time, a detector with such characteristics will be suitable for almost any acoustic experiments throughout its operational frequency range. Thus, in order to prove high performance of our setup, we tested the system by measuring the temperature dependences of the speed and attenuation of longitudinal and transverse ultrasound waves in a relaxor ferroelectric KTa$_{0.92}$Nb$_{0.08}$O$_{3}$ (KTN) single crystal in the vicinity of its ferroelectric transitions.
  
\begin{figure}[th]
\includegraphics[width=1\columnwidth]{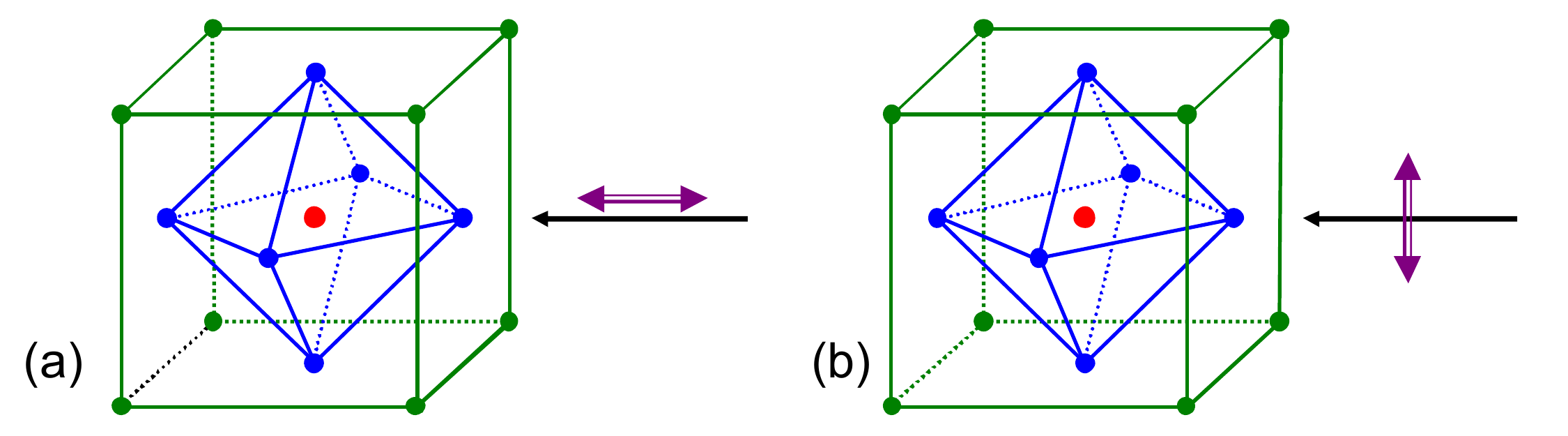}
\caption{KTN structure and orientation of (a) longitudinal ($100|100$), and (b) shear ($100|010$) ultrasound probing pulse with respect to the KTN lattice. The black arrows show the ultrasound propagation direction. The purple double arrows show the ultrasound polarization. The green color shows potassium cubic cages with K$^{+}$ ions at the corners; blue color shows oxygen octahedra with O$^{2-}$ ions at the faces of potassium cubes; the position of the central red ion is occupied by Ta$^{5+}$ or Nb$^{5+}$.}
\label{Fig6}
\end{figure}

KTN possesses the perovskite structure (Fig. \ref{Fig6}) and at room temperature this material is a paraelectric with cubic lattice symmetry (KTN phase diagram is shown in Ref. \onlinecite{Rytz}). Upon cooling, the crystal of the above composition undergoes a sequence of ferroelectric transitions to tetragonal, to orthorhombic, and to rhombohedral lattice symmetry at $T_{c1}=90$~K, at $T_{c2}=84$~K, and at $T_{c3}=76$~K, respectively. 
The crystal used for our tests was (100)-cut with the dimensions $4.82\times5.05\times9.08$~mm${^3}$ and
these transitions we explored by probing them with longitudinal ($100|100$) and shear ($100|010$) ultrasound waves, as shown in Figure \ref{Fig6} (a) and (b), respectively. The use of temperature as an external parameter allowed us to pass through all three transitions.

\begin{figure}[th]
\includegraphics[width=1\columnwidth]{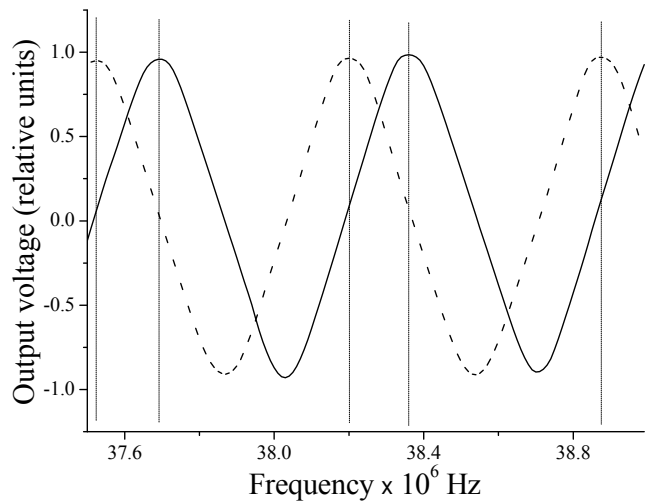}
\caption{Signals of phase detectors as a function of phase shift achieved by varying the probe frequency. The outputs of the phase detectors 1 and 2, see Fig. \ref{fig2}, are shown with solid and dashed lines, respectively.}
\label{Fig7}
\end{figure}

The ultrasound signal was induced in the crystal by means of LiNbO$_{3}$ longitudinal or shear-cut transducers from Boston Piezo-Optics, Inc.\cite{Boston} at their main resonance frequencies of 45.2 and 37.85~MHz, respectively.  These transducers were attached to the optical-quality polished faces of the sample using Stycast-1266 epoxy.\cite{Stycast} The longitudinal wave measurements were performed along the 9.08~mm-long side of the crystal, the transverse wave measurements were done along the 5.05~mm-long side. For low-temperature measurements the sample was placed in a liquid nitrogen cryostat. The temperature was monitored with a 100~$\Omega$ Pt temperature sensor from Sensing Devices, Inc.\cite{Sensing}

\begin{figure}[th]
\includegraphics[width=1\columnwidth]{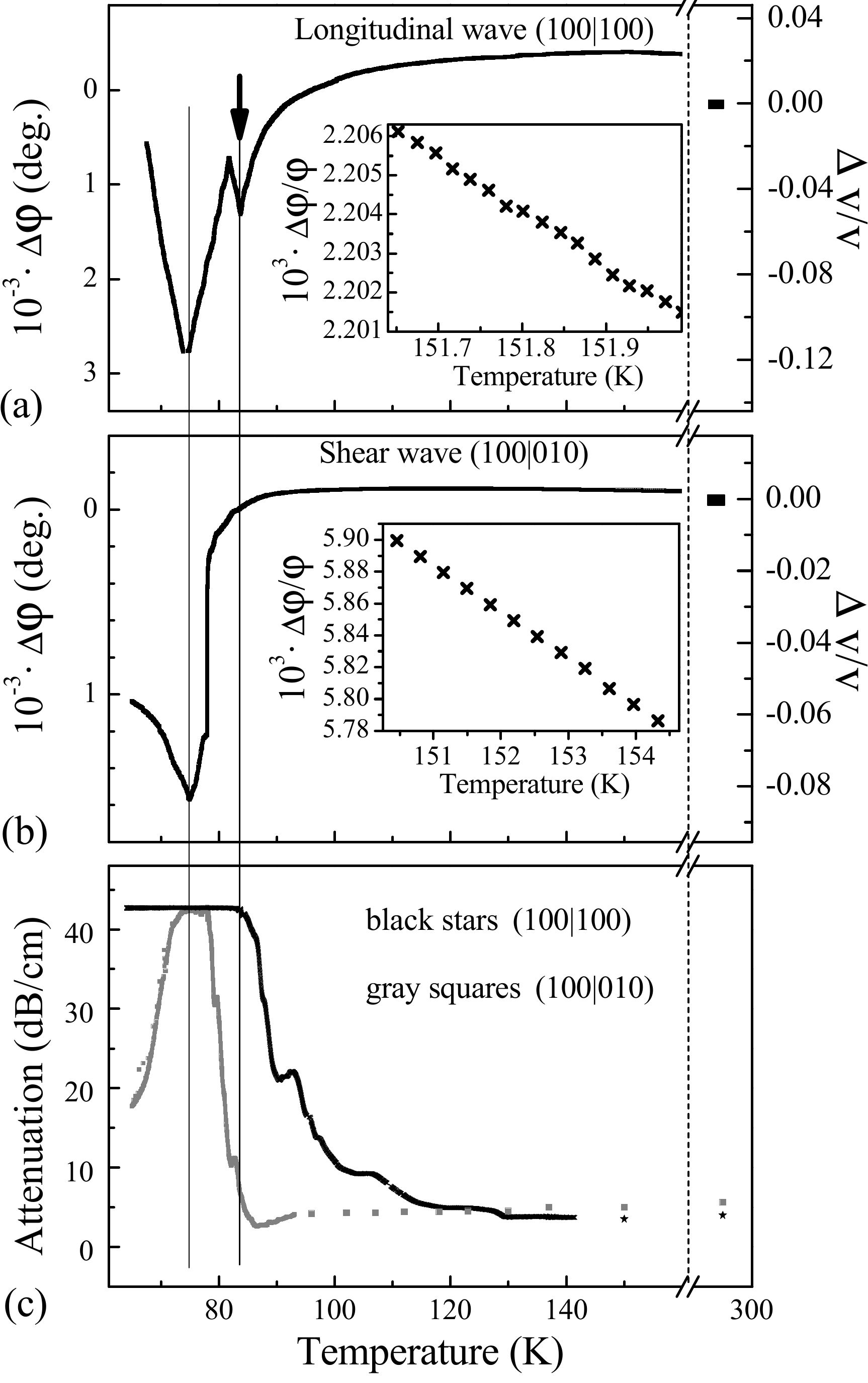}
\caption{ (a) Temperature dependences of the phase shift $\Delta\varphi$ (left axis) and the relative change of the speed (right axis) for longitudinal ultrasound wave in KTN. (b) The same for the transverse wave. The measured data are shown with respect to the room temperature values. The latter is represented by a filled rectangular symbol at 295 K.
(c) Temperature dependence of the attenuation of these waves. 
Insets in (a) and (b) demonstrate high signal-to-noise ratio that provides high sensitivity of the measurements. The arrow in (a) points out a new, but theoretically expected, feature in the longitudinal sound in the vicinity of the tetragonal-to-orthorhombic phase transition.}
\label{Fig8}
\end{figure}

Figure \ref{Fig7} demonstrates the quality of the phase output characteristics of the AD8302 microchips used in the instrument. The phase variation in the sample is related to the signal frequency as $\varphi_0 = 2\pi(2n-1) Lf/v$, where $n$ is the echo number, $L$ is the sample length, $f$ and $v$ are the frequency and the speed of the ultrasound, respectively. This allows for modeling the change of phase of the received signal by varying the frequency of the probing signal sent to the sample. Thus, the transverse wave frequency change from 37.5~MHz to 38.9~MHz corresponds to the phase shift difference by more than $4\pi$. One can see that the output voltages are proportional to the signal phase difference everywhere except for the small regions where the relative phase shift between the received and reference signals was an integer multiplier of $\pi$.  

The measured temperature dependences for longitudinal ($100|100$) and shear ($100|010$) waves are shown in Figure~\ref{Fig8}. 
In panels (a) and (b) we show the change of the signal phase $\Delta\varphi$ with respect to its value at room temperature, as measured by the detector (left axis). Notice, that at the transitions the phase changes by several thousands degrees.

The absolute ultrasound speed at room temperature was determined by a modified pulse overlap technique\cite{overlap, comment} to be equal to 7360$\pm10$~m/s and 3980$\pm10$~m/s for the longitudinal and shear waves, respectively. Therefore, for the first ultrasound echo, the total phase change in the sample $\varphi_0$ in both configurations was about $20000^{\circ}$. These measurements allowed us to estimate the relative phase and the ultrasound speed change, as presented on the right axis in Figure~\ref{Fig8} (a) and (b). Note that these estimates were made under the  assumption that the variation of the sound speed was the only reason for the phase change. A more careful analysis would require accounting for the temperature variations in both the sound and the RF signal paths for which separate experiments were needed. In particular, the sample thermal expansion coefficient would need to be known.

Insets in panels (a) and (b) show expanded fragments of the phase curves demonstrating the high signal-to-noise ratio of the measurements corresponding to a sensitivity to small relative changes in the signal phase on the level of $\sim10^{-6}$, as measured at the first echo (also see Appendix B). However, the accuracy and precision of AD8302 are reduced by a factor of 10 when the phase shift between the input signals is about 45$^{\circ}$. 

The measured temperature dependences of the ultrasound attenuation are presented in panel (c) of Figure~\ref{Fig8}. At high temperature, i.e., far from the transitions, the signal decay was small and changed slowly. In this region, the ultrasound attenuation value at any particular temperature was determined from the exponential decay rate of the ultrasound, i.e. calculated using the amplitudes of all observed echoes.  These results are presented in Figure \ref{Fig8}(c) by points. On approach to the ferroelectric transitions, when the change of attenuation becomes steeper, its value was determined by tracing the gain output of AD8302 of the first (and the only detected) echo. In this way, the relative measurements of the ultrasound attenuation were performed and then recalculated into an absolute value of attenuation. These results are shown in Figure \ref{Fig8}(c) with solid lines. The level of the noise in the attenuation channel determines the sensitivity of the attenuation measurements $\sim$0.05~dB. The accuracy of the attenuation measurements is about 0.5~dB when the attenuation value is above 20 dB.

All of the measured dependences agree with those obtained previously on similar KTN crystals\cite{knauss, svitelskiy}. 
However, one can see that the broad dynamic range and high sensitivity of the AD8302 microchip allowed for recording the ultrasound signal throughout the whole sequence of ferroelectric transitions with a good signal-to-noise ratio, as shown in Figures \ref{Fig8} (a-c). To the best of our knowledge, acoustic measurements through these transitions have never been reported before. We, and other authors, used to terminate the acoustic experiments in the vicinity of the first of the transitions, where the high value of the sound attenuation made further measurements impossible. Thus, the results of the above tests prove the exceptional characteristics of the reported setup. Also the high sensitivity of this new instrument allowed for a resolution of the theoretically predicted hardening of the longitudinal wave upon completion of the transformation of the crystal to the tetragonal lattice structure \cite{Levanyuk} (see the marked-with-arrow speed increase just below T = 84 K in Fig. \ref{Fig8} (a)).

\begin{figure*}[th]
\includegraphics[width=2\columnwidth]{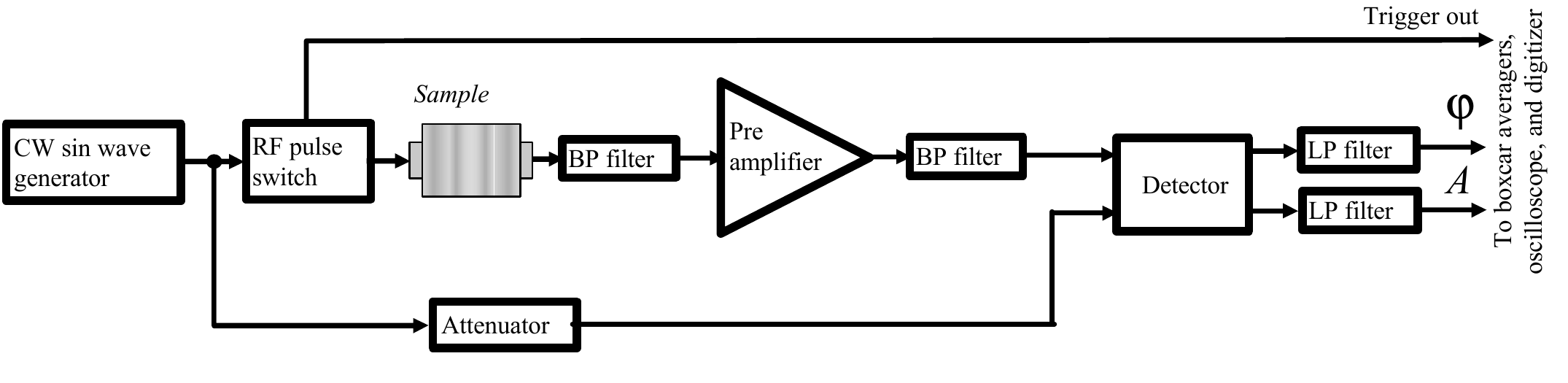}
\caption{Block diagram of a simplified version of our setup: only one detector AD8302 is used (compare with Figure \ref{fig2}).}
\label{simplified}
\end{figure*}

\section{Conclusion}

In this paper, we report on a superior quality ultrasound detector made with modern mass-produced analog microchips for RF electronics. The utilization of a broadband AD8302 gain and phase detector microchip with precisely balanced inputs allowed for an exclusion of the need for superheterodyne circuitry and for treating the signal directly at the measurement frequency, which in turn, significantly simplified the entire instrument. A continuity of measurements over the large range of change of the sound speed in the sample is achieved by parallel connection of two AD8302 detectors. 

The approach presented here of utilizing the in-phase (I) and quadrature (Q) reference signals is broadly used in the superheterodyne I/Q mixer circuits. However, by utilizing the modern integrated microchips and the direct conversion technique, we propose a new elegant and simple solution for building a detector for ultrasonic measurements. It is important to note that our instrument makes ultrasound experiments accessible not only to a narrow circle of acoustics experts, but also to a broad range of non-specialist researchers.

The performance of the instrument was tested in the laboratory environment  by studying the strong temperature dependences of the ultrasound speed and attenuation in a KTa$_{0.92}$Nb$_{0.08}$O$_{3}$ crystal undergoing a sequence of ferroelectric transitions. We show that due to its high sensitivity and large dynamic range the detector facilitates experiments where high precision at a large change of the signal amplitude and phase (i.e., ultrasound attenuation and speed) is required. Despite its simplicity, the instrument demonstrates highly competitive performance. Thus, the instrument allows us to perform measurements at a previously unavailable temperature range thereby revealing a theoretically expected, but previously unobservable feature in the speed of the longitudinal sound wave in KTN.

\section{Acknowledgements}

This work was partially supported by the Colgate University Research Council and the Department of Physics and Astronomy at Colgate University. The National High Magnetic Field Laboratory (for A. S.) is supported by National Science Foundation Cooperative Agreement No. DMR-1157490, and the State of Florida. Research at the Oak Ridge National Laboratory (for L. A. B.) is sponsored by the US Department of Energy, Office of Science, Basic Energy Sciences, Materials Sciences, and Engineering Division. J. G. and O. S. are thankful to Professors K. Segall, J. Amato, E. Galvez, B. Parks, and R. Metzler for various help throughout the whole project, and to Professor C. H. Holbrow for the critical review of this work. The authors are grateful to Mr. C. Augusta for providing technical support on AD8302, to Analog Devices, Inc. for donating microchips and to Sensing Devices, Inc. for donating temperature sensors.

\section*{Appendix A: Simplified Setup}

In Figure \ref{simplified} we present a block diagram of a simplified version of our setup that includes only one detector AD8302. Such a setup was used for measurements where the observed change of the ultrasound speed was small. If during an experiment the signal phase change caused by the ultrasound velocity variation does not exceed $\sim140^{\circ}$ (i.e., the phase resides within the linear segment of the AD8302 detector), this simplified setup can be used without any special precautions.

\section*{Appendix B: Stability and sensitivity of the AD8302}

\begin{figure*}[th]
\includegraphics[width=2\columnwidth]{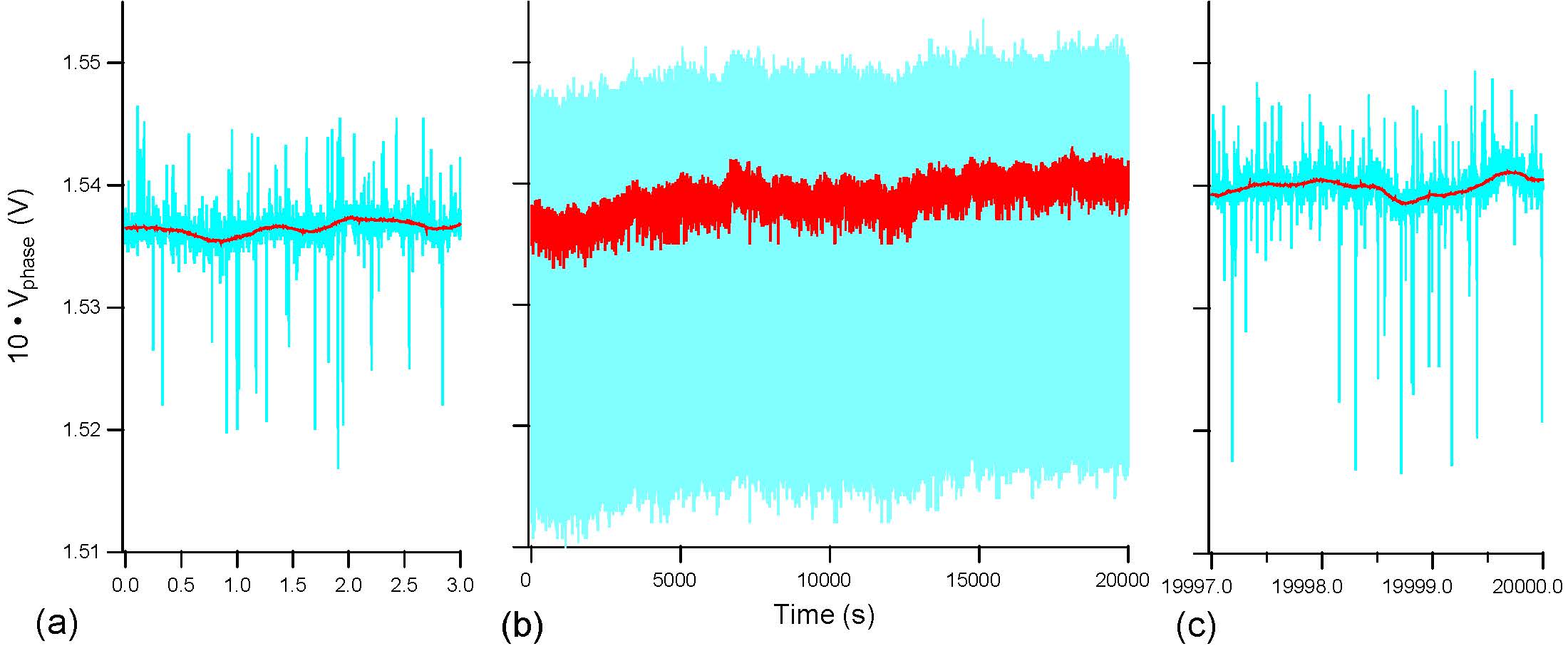}
\caption{(Color online) Illustration of the long-term stability of the instrument: time dependence of the AD8302 phase output. (a) The first three seconds of the run. (b) The entire run. (c) The last three seconds of the run. The vertical scale is the same in all three (a)-(c) panels. The light-blue line shows the as-measured signal, the dark-red shows the signal averaged for 0.3 seconds.}
\label{stabilityPhase}
\end{figure*}

\begin{figure*}[th]
\includegraphics[width=2\columnwidth]{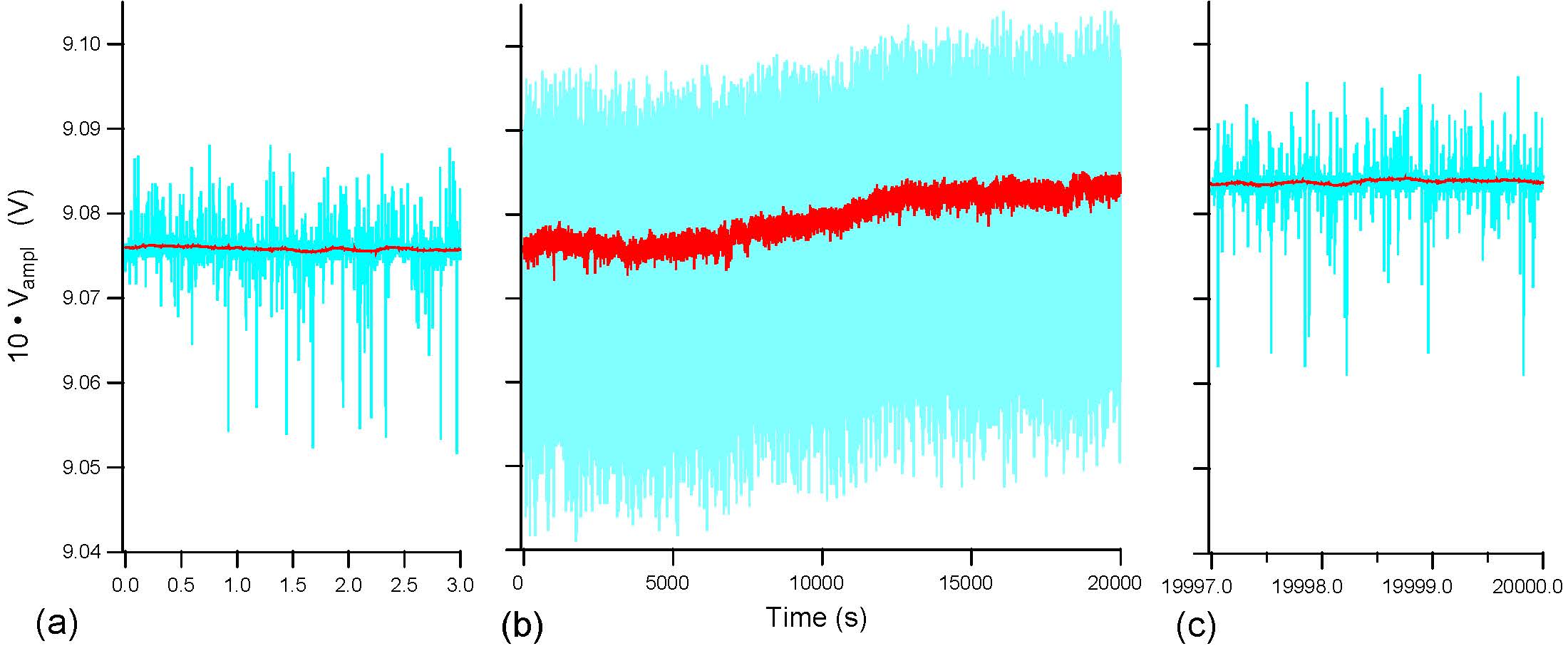}
\caption{(Color online) Illustration of the long-term stability of the instrument: time dependence of the AD8302 amplitude output. (a) The first three seconds of the run. (b) The entire run. (c) The last three seconds of the run. The vertical scale is the same in all three (a)-(c) panels. The light-blue line shows the as-measured signal, the dark-red shows the signal averaged for 0.3 seconds.}
\label{stabilityAmp}
\end{figure*}

In order to check the long-term stability of our instrument, we have operated it while keeping all of the experimental conditions constant over a period of 5-6 hours, which is about 2 times longer than the actual time of our experiments described in Part \ref{partTest}. As in the actual experiments, in this test we also used a boxcar averager SR250, which provided a gain of 10 to the AD8302 output voltages. The repetition rate of the ultrasound excitation pulse was 10 kHz, and the boxcar averaged the AD8302 voltage over 3000 samples. Therefore, the boxcar time constant was about 0.3 s. The boxcar output voltage was acquired by an analog-to-digital converter, which was set to make 1000 measurements per second. The time dependences of the boxcar output voltages (i.e., of AD8302 output voltages multiplied by factor of 10) are presented in Figures \ref{stabilityPhase} and \ref{stabilityAmp}. In both Figures, panel (a) shows the output phase voltage of one of the detectors during the first three seconds of the experiment; panel (b) demonstrates how the signal changed over the entire 20,000-second interval; and panel (c) shows the last three seconds of the run. In all panels, the light-blue line shows the as-measured signal, and the dark-red line shows the signal averaged in a computer over 300 samples, which corresponds to a boxcar time constant of 0.3 s. 

The difference shown in Figure \ref{stabilityPhase}(b) between the boxcar output voltage at the beginning and at the end of the measurement is about 4 mV, which corresponds to a 0.4 mV change in the AD8302 phase output voltage and to the phase shift of $0.04^{\circ}$ (according to the AD8302 specification, the phase measurement scaling is 10 mV/deg). One can also see that the noise magnitude of the as-measured signal in panels (a) and (c) in Figure \ref{stabilityPhase} is 2 mV, i.e., about  $0.02^{\circ}$. This establishes the sensitivity to the phase change (which we define as a change that is three times larger than the noise level) of about $0.06^{\circ}$. (When we estimate the noise magnitude, we exclude the spikes present on the curves. Such a width of 2 mV includes about 95\% of the signal and represents the 2$\sigma$ interval). The stability of our instrument meets well the requirements to the detectors used for the measurements of an absolute sound speed,\cite{overlap, comment} which usually take not more than few minutes. 

Similarly, in Figure \ref{stabilityAmp}(b) the boxcar signal change during 5 hours is about 7 mV. This corresponds to a 0.7 mV change in the AD8302 amplitude output, i.e., to a 0.02 dB change in the signal amplitude. However, the 2$\sigma$ intervals of the as measured signal in Figure \ref{stabilityAmp}(a) and Figure \ref{stabilityAmp}(c) is about1.5 mV; this gives the sensitivity to the amplitude change at a level of about 0.05 dB.

The parameters of the AD8302 detector presented above are significantly better than the specifications provided in the AD8302 datasheet \cite{Analog}. However, it should be taken into account that (according to information obtained from the technical support of  Analog Devices, Inc.) in the plots presented in the datasheet in Figures TPC 26 -- TPC 29 the output of the AD8302 was averaged only over about 16 measurements. Also note that the manufacturer's datasheet includes the data on the variation of the parameters between different microchips. Thus, for example,  Figures TPC 30 -- TPC 32 in the datasheet represent results obtained on a set of about 17000 microchips. 

Therefore, the high stability and sensitivity (phase and amplitude discrimination) of our detector is due to the high quality of these parameters in each single microchip and by virtue of signal processing by the boxcar averager (and the computer) used in our experiments and tests. At the same time, note that the nonlinearity in the phase measurements increases as the phase difference between the input signals is close to an integer multiplied by $\pi$. Thus, even in the two-detector configuration (see Figure \ref{fig2}) the accuracy of the phase measurements might be as low as $0.5^{\circ}$. Also the nonlinearity (i.e., the accuracy) in the amplitude measurements might be as high as 0.5 dB at a large difference in the levels of the input signals.

\begin{thebibliography}{99}
\bibitem{comment} While a precise measurement of the absolute value of sound velocity is out of the scope of this project, it is worth noting that the developed ultrasonic techniques (such as the methods of overlap, superposition, etc.) allow for canceling the errors, which are due to the difference of the RF signal paths. In these techniques, the absolute measurements are converted into measurements of a relative phase shift. Therefore, our detector, in general, allows for precise absolute measurements of the sound velocity, but at the same time, the accuracy of the obtained velocity value will depend on the sample length measurement, sample wedging, the bond thickness, etc. Methods for absolute velocity measurements have been reviewed in many publications. See, for example, Ref. \onlinecite{overlap} and Refs. therein.
\end{thebibliography}
\end{document}